\documentclass{elsart}               

\usepackage{amsmath}
\usepackage{amssymb}
\usepackage{bm}
\input colordvi

\begin{document}

\begin{frontmatter}

\title{Decelerating and accelerating back-reaction of vacuum to the Universe expansion.
} 
\author[Cherkas]{S. L. Cherkas}\ead{cherkas@inp.minsk.by},    
\author[Kalashnikov]{V. L. Kalashnikov}\ead{v.kalashnikov@tuwien.ac.at}
\address[Cherkas]{ Institute for Nuclear Problems, Bobruiskaya 11, Minsk 220050,
Belarus}  
\address[Kalashnikov]{Institut f\"{u}r Photonik, Technische Universit\"{a}t Wien,\\
Gusshausstrasse 27/387, Vienna A-1040, Austria}             


\begin{abstract}                          
Back-reaction of the massless scalar field vacuum to the Universe
expansion  is considered. Automatic renormalization procedure
based on the equations of motion instead of the Friedman equation
is used to avoid the cosmological constant problem. It is found,
that the vacuum tends to decelerate an expansion of Universe if
the conformal acceleration of Universe is equal to zero. In
contract, the vacuum acts as the true cosmological constant in
Universe expanding with the present day acceleration. Estimation
for third derivative of the Universe scale factor with respect to
conformal time is presented.

\end{abstract}

\end{frontmatter}

\section{Introduction}
It is generally well-known, that the vacuum cosmological constant
calculated from the momentum cutting at the Plank level is much
greater than the observed one (this problem has a long history,
see \cite{weinberg,carr,pd}, for instance). Thus, one needs to
renormalize mean value of the vacuum energy of Universe or, more
precisely, to renormalize the Friedman equation in order to
provide a reasonable value of the cosmological constant.

On the one hand, the problem stimulates a search of the new kinds
of matter, which would emulate the cosmological constant
\cite{ell,stein,arm,cald,gibb,guen}. On the other hand, the issue
is a deeper understanding of the vacuum at non-stationary and
curved space-times. Considerable successes in the renormalization
programm, as applied to the cosmological constant problem was
reached both earlier \cite{birrel,od} and at present
\cite{on,par}. However, one has note that the aim of
renormalizations (treated in the QED style) is to exclude the
ultraviolet (UV) cutoff from final result. While after advance in
the string theory, it is possible to guess the UV cutoff at the
Planck level. Thus, renormalizations involving the UV cutoff and
providing a reasonable value of the cosmological constant are
acceptable in principle.

It is already mentioned, that the UV cutoff for the direct vacuum
energy counting results in the cosmological constant value, which
is enormously large \cite{weinberg,carr,pd}. However, as it was
shown in Ref. \cite{pad}, the cosmological constant can be
proportional to the squared Hubble constant if one uses the square
root from the dispersion of the vacuum energy fluctuations instead
of the vacuum energy itself. Calculations of these fluctuations in
the different ways and with the different argumentation have been
presented \cite{zel,pad0,bomb,sred,brust,ver,yar}. Nevertheless,
identification of the cosmological constant source with the vacuum
energy fluctuations is still ungrounded at the fundamental level.

As it will be shown below, the true value of the cosmological
constant can be obtained by consideration of another quantity,
this is the mean value of difference of the potential and kinetic
energies of field oscillators in the vacuum state. Under the UV
cutoff, this quantity turns out to be much smaller than the sum of
these energies appearing in an ordinary reasoning concerning the
cosmological constant problem. The corresponding  renormalization
procedure is based on the quasi-classical second-order equation of
motion for the Universe scale factor instead of the Friedman
equation. This procedure can be considered as a quasi-classical
limit of the quantization scheme for equations of motion suggested
in \cite{prep1,prep2} to describe a dynamics of the quantum
Universe, which violates the Hamiltonian constraint $H=0$ (i.e.
the Friedman equation).

As a result of such renormalization procedure, it is possible to
connect the observed value of the Universe acceleration with the
present day Hubble constant and with the vacuum state of a
massless scalar field.

In addition, this allows predicting the third derivative of the
Universe scale factor, which can be observable within the next few
years.

\section{Classical dynamics}

Let us begin with the Einstein action for a single-component real
scalar field:
\begin{equation}
S=\frac{1}{16\pi G}\int d^4 x\sqrt{-g}R+\int d^4 x\sqrt-g[\frac{1}{2}%
(\partial_\mu\phi)^2-m^2\phi^2), \label{deistv}
\end{equation}
and consider a homogeneous, isotropic and flat Universe with the
metric:
\begin{equation}
ds^2=a^2(\eta)(N^2(\eta)d\eta^2- {d\bm r}^2).
\end{equation}
\noindent Taking into account an inhomogeneity of the scalar
field, one can write:
\begin{equation}
S=V\int \left( -\frac{1}{2}M_p^2\,\frac{(a^\prime)^2}{N}
+\frac{1}{2}\sum_{\bm k}\frac{ a^2\,\phi_{\bm k}^\prime\phi_{-{\bm
k}}^\prime}{N}- a^2 N k^2\phi_{\bm k}\phi_{-{\bm k}} -a^4 N
m^2\phi_{\bm k}\phi_{-{\bm k}}\right)d\eta, \label{action}
\end{equation}
where  $M_p=\sqrt{\frac{3}{4 \pi G}}$ is the Plank mass and $V$ is
the three-dimensional volume. Further, we shall fix the gauge by
the choice of $N=1$. Thus, we shall use the conformal time $\eta$
\cite{lan}, which is connected with the cosmic time $t$ by
relation $dt =a(\eta) d\eta$. Prime means the differentiation over
$\eta$, and $\phi_k$ is the Fourier-transform of $\phi(\bm
r)=\sum_{\bm k} \phi_{\bm k} e^{i {\bm k}\bm r}$. Redefinitions $
a^2 \rightarrow a^2/V $ and $m^2\rightarrow m^2 V$ allow omitting
the volume $V$ in intermediate calculations. Corresponding
Hamiltonian is
\begin{equation}
H=-\frac{p_a^2}{2 M_p^2}+\sum_{\bm k} \frac{\pi_{\bm k}\pi_{-{\bm
k}}}{2 a^2}+\frac{1}{2} a^2 k^2 \phi_{\bm k}\phi_{-{\bm k}}
+\frac{1}{2} a^4 m^2 \phi_{\bm k}\phi_{-{\bm k}},
\label{ham0}
\end{equation}
where momentums are connected with the velocities by the relations
$\phi^\prime_{\bm k}=\frac{\pi_{-{\bm k}}}{a^2}$ and
$a^\prime=\frac{{p_a}}{M_p^2}$. In terms of velocities the
Hamiltonian is:
\begin{equation}
H=-\frac{1}{2}M_p^2(a^\prime)^2+\frac{1}{2}\sum_{\bm k}
a^2\phi_{\bm k}^\prime\phi_{-{\bm k}}^\prime +k^2 a^2 \phi_{\bm
k}\phi_{-{\bm k}} +a^4 m^2 \phi_{\bm k}\phi_{-{\bm k}}.
\label{ham}
\end{equation}
Equations of motion can be obtained by varying the action
(\ref{action}) or from the Hamiltonian (\ref{ham0}) by the means
of the Poisson brackets \cite{prep2}:
\begin{eqnarray}
\phi^{\prime\prime}_{\bm k}+(k^2+a^2 m^2)\phi_{\bm k}+2\frac{
a^\prime}{a}{ \phi^\prime}_{\bm k}=0,\\
M_p^2\,a^{\prime\prime}+\frac{1}{a}\sum_{\bm k}
a^2\phi^\prime_{\bm k}\phi^\prime_{-{\bm k}} -a^2 k^2\phi_{\bm
k}\phi_{-{\bm k}} -2 a^4 m^2 \phi_{\bm k}\phi_{-{\bm k}}=0.
\label{sys}
\end{eqnarray}
It should be noted, that, besides, there is the Friedman equation
$H=0$ arising from variation of the action (\ref{action}) over
$N$. Thus, at a classical level one may add to Eq. (\ref{sys}) the
equation $H=0$ multiplied by some coefficient (here $H$ is given
by Eq. (\ref{ham})). Our postulate is that at a quantum level, one
has to use the equation of motion for the Universe scale factor
containing exact difference of the potential and kinetic energies.
In the general case of a massive scalar field, one cannot build
such an equation. However, it is possible for a massless scalar
field. Moreover, in the conformal time, it is not necessary to add
additional terms to Eq.(\ref{sys}) in order to provide a mutual
difference of the potential and kinetic energies of a scalar
field, since it already contains their exact difference in the
case of $m=0$.

\section{Quasiclassical picture}

Let us consider Universe scale factor as a classical quantity in
the equations:
\begin{eqnarray}
{\hat \phi}^{\prime\prime}_{\bm k}+k^2{\hat\phi}_{\bm k}+2\frac{
a^\prime}{a}{ {\hat\phi}^\prime}_{\bm k}=0,\label{sys1a}\\
M_p^2\,a^{\prime\prime}+\frac{1}{a}\sum_{\bm k}
a^2<\hat{\phi}^\prime_{\bm k}{\hat \phi}^\prime_{-{\bm k}}> -a^2
k^2<{\hat \phi}_{\bm k}{{\hat \phi}}_{-{\bm k}}> =0, \label{sys1}
\end{eqnarray}
and use the mean values   $<{\hat\phi}^\prime_{\bm k}{\hat
\phi}_{-{\bm k}}^\prime>$ and $<{\hat \phi}_{\bm
k}{\hat\phi}_{-{\bm k}}>$ in the second equation. According to Eq.
(\ref{sys1a}), a quantum massless scalar field evolves at a
classical background. For the stationary space-time $a=const$, the
`` masses'' of the field oscillators (which are $\sim a^2$) are
constant. According to the virial theorem, in the ground state,
the mean value of the potential energy is equal to that of the
kinetic one. Consequently, there exists the natural solution
$a=const$ of Eqs. (\ref{sys1a},\ref{sys1}) with the scalar field
oscillators in a ground state.

Quantization of the scalar field \cite{birrel}
\begin{equation}
\hat \phi(r)=\sum_{\bm k} \hat {\mbox{a}}^+_{\bm
k}\psi_{k}^*(\eta)e^{-i{\bm k}\bm r}+\hat {\mbox{a}}_{\bm k}
\psi_{k}(\eta)e^{i {\bm k}\bm r}
\end{equation}
leads to the operators of creation and annihilation with the
commutation rules
 $[{\hat{\mbox{a}}}_{\bm k},\, {\hat{\mbox{a}}}^+_{\bm k}]=1$. The complex functions
 $\psi_k(\eta)$ are
 $\psi_k=\frac{1}{a\sqrt{2 k}}e^{-ik\eta }$ for $a=const$. In the general case
 they are solutions of the equation
\begin{equation}
\psi^{\prime\prime}_k+k^2\psi_k+2\frac{ a^\prime}{a}{
\psi^\prime}_k=0 \label{osc}
\end{equation}
 and satisfy to the relation \cite{birrel,kim}:
\begin{equation}
a^2(\eta)(\psi_k \,{\psi_k^\prime}^*-\psi_k^*\,\psi_k^\prime)=i.
\label{wronskian}
\end{equation}

Fourier components of the scalar field can be written in the form
\begin{equation}
\hat \phi_{\bm k}=\frac{1}{V}\int_V\hat \phi(\bm r)e^{-i{\bm k}\bm
r}d^3\bm r=\hat{\mbox{a}}_{-{\bm
k}}^+\psi_k^*(\eta)+\hat{\mbox{a}}_{{\bm k}}\psi_k(\eta).
\end{equation}
The corresponding momentums are:
\begin{equation}
\hat \pi_{\bm k}=a^2(\eta)\hat \phi_{-{\bm
k}}^\prime=a^2(\hat{\mbox{a}}_{{\bm k}}^+{\psi_k^\prime}^*+\hat
{\mbox{a}}_{-{\bm k}}\psi_k^\prime).
\end{equation}
This gives the ordinary commutation relations  $[\hat \pi_{\bm k},
\hat \phi_{{\bm k}}]=-i$.

Let us assume that the field is in the vacuum state, then
\begin{equation}
a^2<0|\hat \phi^\prime_{\bm k}\hat\phi^\prime_{-{\bm k}}|0>-
k^2a^2 <0|\hat\phi_{\bm k}\hat\phi_{-{\bm k}}|0>=
a^2({\psi_k^\prime}^{*}\psi_k^\prime-k^2\psi_k^*
\psi_k)=\partial_\eta \sigma_k, \label{15}
\end{equation}
where
$\sigma_k=\frac{1}{2}a^2(\eta)({\psi_k^\prime}^*\psi_k+\psi_k^\prime\psi_k^*)$.

It should be noted that, besides the considered Heisenberg picture
of the field oscillators, the Schr\"{o}dinger picture exists, too.
In this picture, the operators $\hat \phi_k$ and $\hat \pi_k$ do
not depend on the time $\eta$, but the states of the field
oscillators do depend (see \cite{kim} and references therein).
Eigenstates of the Ermakov-Lewis invariant can be build in such a
representation \cite{ermakov,lewis}, and the lowest eigenstate is
the vacuum state.

Now we turn to consideration of the vacuum back-reaction to the
Universe scale factor evolution. The simplest way of that is to
 calculate quantity  $\sum_k \sigma^\prime_k$ for the different
 kinds of dependence
 $a$ on $\eta$ (the Universe expansion law).  For instance, let us take $a(\eta)$ in the form
\begin{equation}
a(\eta)=a_0(1+\alpha\, \mathcal H \,\eta)^{1/\alpha}, \label{acon}
\end{equation}
to satisfy $a(0) \equiv a_0$ and $\mathcal H \equiv a^\prime/a$ at
$\eta=0$, which will be identified with the present moment of
time. The conformal Hubble parameter $\mathcal H$ is connected
with the ordinary one as $H=\frac{{\dot a}}{a}=\frac{1}{a}\mathcal
H=\frac{a^\prime}{a^2}$, where dot is the derivative over the
cosmic time $t$.  Two values of $\alpha$ parameter: $\alpha=1$ and
$\alpha=-1/2$ are of particular interest.

In the first case, the conformal acceleration of Universe
$a^{\prime\prime}$ equals to zero. Dependence of the scale factor
on the cosmic time has the form: $a(t)=a_0\sqrt{1+\frac{2\mathcal
H t}{a_0}}$.

In the second case, the parameter $\alpha=-1/2$ gives the
conformal acceleration
$\frac{a^{\prime\prime}a}{{a^\prime}^2}=3/2$, which corresponds to
$\frac{\ddot a\, a}{{\dot
a}^2}=\frac{a^{\prime\prime}a}{{a^\prime}^2}-1=1/2$ in the cosmic
time units and is close to the observational data $\frac{\ddot a\,
a}{{\dot a}^2} \approx 0.55$ \cite{rich}. Dependence of the scale
factor on the cosmic time reads as $a(t)=a_0\left(
1+\frac{\mathcal H t}{2 a_0}\right)^2$ for $\alpha=-1/2$.

For dependence $a(\eta)$ in the form of Eq. (\ref{acon}), Eqs.
(\ref{osc}),(\ref{wronskian}) are satisfied by
\begin{equation}
\psi_k(\eta)=\frac{-i\sqrt{\pi}}{2 a_0\sqrt{\mathcal H \alpha}}
\exp\left(\frac{ik}{\mathcal H \alpha}+\frac{i \pi}{2
\alpha}\right) (1+\alpha \mathcal H
\eta)^{1/2-1/\alpha}H^{(2)}_{1/2-1/\alpha}\left(\frac{k}{\mathcal
H \alpha}+k \eta\right),
\end{equation}
where $H_n^{(2)}(x)=J_n(x)-iY_n(x)$ is the Hankel function of the
second kind. When $\mathcal H\rightarrow 0$, the function
$\psi_k(\eta)$ tends to $\frac{1}{a_0 \sqrt{2 k}}e^{-ik \eta}$.

For the case of $\alpha=1$ calculation of  $\sigma_k$ gives
\begin{equation}
\sigma_k=-\frac{\mathcal H}{2k(1+\mathcal H \eta)}.
\end{equation}

According to the relation
\begin{equation}
a^{\prime\prime} a=-\frac{1}{M_p^2V}\sum_k \sigma^{\prime}_k
\label{aa}
\end{equation}
(dependence on the volume $V$ is restored), which follows from
Eqs. (\ref{sys1},\ref{15}), the vacuum back-reaction is given by

\begin{equation}
-\frac{1}{M_p^2 V}\sum_{\bm k} \sigma^{\prime}_k=
-\frac{1}{(2\pi)^3M_p^2}\int_{k_{min}}^{k_{max}}\sigma^{\prime}_k\,
d^3 {\bm k}= -\frac{{\mathcal H}^2(k_{max}^2-k_{min}^2)}{8 \pi^2
M_p^2 (1+\mathcal H \eta)^2}.
\end{equation}

One can see, that the vacuum causes a deceleration with the
negative acceleration parameter
$\frac{a^{\prime\prime}a}{{a^{\prime}}^2}\approx-\frac{k_{max}^2}{8\pi^2
a_0^2
 M_p^2}$ at the present time $\eta=0$. Natural cutting of the physical momentums
$k_{max}/a_0$ at the level of $k_{max}/a_0\sim M_p\, \pi \sqrt{8}$
results in $\frac{a^{\prime\prime}a}{{a^{\prime}}^2}\sim -1$. Let
us remind that the starting assumption was an absence of
acceleration $a^{\prime\prime}=0$, given by Eq. (\ref{acon}) with
$\alpha=1$. Thus, Universe with the zero conformal acceleration
could exist only in a presence (besides vacuum) of some substance
(like Einstein's cosmological constant) compensating the
decelerating action of vacuum.

Let us turn to the case of Universe expanding with the observed
acceleration corresponding to $\alpha=-1/2$. Calculation of
$\sigma_k$ leads to
\begin{eqnarray}
\sigma_k=-\frac{3\,\mathcal H^3}{2\,k^3\,{\left( 2 -\mathcal
H\,\eta \right) }^3} - \frac{\mathcal H}{k\,\left( 2 - \mathcal
H\,\eta \right) },
\end{eqnarray}
where $\eta \in (-\infty, 2/\mathcal H)$.
 Note, that these $\sigma_k$ satisfy to Eq.
(\ref{single})  derived in Appendix, when the appropriate
$a(\eta)$ is considered.

Finally, for the vacuum back-reaction one can obtain:
\begin{eqnarray}
-\frac{1}{M_p^2 V}\sum_{\bm k} \sigma^{\prime}_k=
-\frac{1}{(2\pi)^3M_p^2}\int_{k_{min}}^{k_{max}}\sigma^{\prime}_k\,
d^3 {\bm k}=
\nonumber\\
\frac{{\mathcal H}^2(k_{max}^2-k_{min}^2)(2-\mathcal H \eta)^2 + 9
{\mathcal H}^4 \ln\left(\frac{k_{max}}{k_{min}}\right)}{4 \pi^2
M_p^2 (2-\mathcal H \eta)^4}.
\end{eqnarray}
This means that at $\eta=0$  acceleration parameter is positive:
\begin{equation} \frac{a^{\prime\prime}a}{{a^{\prime}}^2}\approx
\frac{k_{max}^2}{16\pi^2 a_0^2  M_p^2}, \label{form}
\end{equation}
and is an order of $\frac{a^{\prime\prime}a}{{a^{\prime}}^2}\sim
3/2$ under cutting $k_{max}/a_0\sim 4\pi\, M_p\, \sqrt{{3}/{2}}$.
This value corresponds to the initial assumptions. Thus, the
observed acceleration of Universe can be caused by a pure vacuum
of the massless scalar field under natural cutting of the
momentums at the Plank level. Note, that influence of lowest
cutting is negligible and $k_{min}$ can be chosen as  $k_{min}
\sim \mathcal H$.

It should be emphasized, that in our consideration the
cosmological constant is derived not from the vacuum energy
dispersion, but from the mean value of difference of the potential
and kinetic energies in the vacuum state of the field oscillators.

  In the near future, measurements
of the third derivative of the Universe scale factor should be
available \cite{star}. It is possible to predict the value of
parameter
$\frac{(a^{\prime\prime}a)^\prime}{a^{\prime\prime}a^{\prime}}=
1+\frac{a^{\prime\prime\prime}a}{a^{\prime\prime}a^{\prime}}$ at
the present time $\eta=0$:
\begin{equation}
\frac{(a^{\prime\prime}a)^\prime}{a^{\prime\prime}a^{\prime}}=
\frac{a}{a^\prime}\frac{(a^{\prime\prime}a)^\prime}{a^{\prime\prime}a}
=\frac{1}{\mathcal H }\frac{\sum_{\bm k}
\sigma^{\prime\prime}_k}{\sum_{\bm k}
\sigma^{\prime}_k}=\frac{4\,\left( {{k_{max}}}^2 -
       {{k_{min}}}^2 \right)  +
    18\,\mathcal H^2\,\ln (\frac{{k_{max}}}
       {{k_{min}}})}{4\,
     \left( {{k_{max}}}^2 -
       {{k_{min}}}^2 \right)  +
    9\,\mathcal H^2\,\ln (\frac{{k_{max}}}
       {{k_{min}}})}\approx 1,
\label{one}
\end{equation}
where Eq. (\ref{aa}) was used. On the other hand, the dependence
of $a$ on $\eta$ given by Eq. (\ref{acon}) for $\alpha=-1/2$ leads
to
\begin{equation}
\frac{(a^{\prime\prime}a)^\prime}{a^{\prime\prime}a^{\prime}}= 3.
\label{two}
\end{equation}

 The difference between Eqs. (\ref{one}) and (\ref{two}) arises
 because, in fact, we used the iteration procedure: i)
preassumption of the Universe expansion given by Eq. (\ref{acon})
allows ii) calculating the vacuum back-reaction. As the next step,
iii) correction of the $a(\eta)$ dependence is required. However,
we restrict ourself to the steps i) and ii), because this allows a
completely analytical consideration. In the general case,
self-consistent system of equations (\ref{a}),(\ref{single}) (see
Appendix) has to be solved.

Above, only massless scalar particles are under consideration,
though one can guess that an analogous consideration can be
extended to include the photon and graviton vacuums. The question
arises: what is with the vacuum of the massive particles
(including inflanton) to do? One can guess, that some
renormalization scheme relevant immediately to the Lagrangian can
be developed for this aim because the Lagrangian always contains
the exact difference of the kinetic and potential energies both
for massless and for massive particles.

It is worth to be noted, that only one kind of the particles has
been taken into account in the estimation (\ref{one}). However, it
still can be applied to the real Universe. This results from its
approximate independence on the upper cutoff and thus its
independence on the variety of the particles.

\section{Conclusion}

It has been shown, that the present acceleration of Universe can
result from the back-reaction of the massless scalar field vacuum
if the momentums cutting lies at the Plank level. The main
assumption is that the renormalization procedure for the vacuum
energy of the massless scalar field is based on the equation of
motion for the Universe scale factor instead of the Friedman
equation. The equation of motion has the form, where the kinetic
energy is subtracted from the potential one. The procedure allows
the rough estimation for the parameter
$\frac{a^{\prime\prime\prime}a}{a^{\prime\prime}a^{\prime}}\sim
0\div 2$, where $a^{\prime\prime\prime}$ is the third derivative
of the Universe scale factor over the conformal time.

\appendix
\section{Self-consistent system of equations for arbitrary state of the scalar field}
It is possible to write down the system of equations for an
arbitrary (not vacuum) state of the field oscillators. Let us
denote mean values $\chi_{\bm k}=<\hat \phi_{\bm k}\hat\phi_{-\bm
k}>$, $\theta_{\bm k}=<\hat \pi_{\bm k}\hat \pi_{-\bm k}>$ and
$\sigma_{\bm k}=\frac{1}{2}<\hat \pi_{-\bm k}\hat\phi_{-\bm
k}+\hat \phi_{\bm k}\hat \pi_{\bm k}>$.

Taking the derivatives and using the relations $\hat\pi_{\bm
k}=a^2\hat \phi_{-\bm k}^\prime$, $\hat \pi_{\bm k}^\prime=(a^2
\hat \phi_{-\bm k}^\prime)^\prime=a^2\hat\phi_{-\bm
k}^{\prime\prime}+2a \,a^\prime\hat \phi_{-\bm k}^\prime=-a^2
k^2\hat \phi_{-\bm k}$
 lead to the equations:
\begin{eqnarray}
\chi^\prime_{\bm k}=\frac{2}{a^2}\sigma_{\bm k},\nonumber\\
\theta_{\bm k}^\prime=-2 a^2 k^2 \sigma_{\bm k},\nonumber\\
\sigma_{\bm k}^\prime=-a^2 k^2\chi_{\bm k}+\frac{\theta_{\bm
k}}{a^2}. \label{mean}
\end{eqnarray}
With the equation for the scale factor:
\begin{equation}
M_p^2a^{\prime\prime}+\frac{1}{a}\sum_{\bm k} \sigma_{\bm
k}^\prime=0, \label{a}
\end{equation}
the above equations give the closed system.

Eqs. (\ref{mean}) have the same form as that for the single
time-dependent oscillator \cite{hez}. There exists the integral of
motion of (\ref{mean}):
\begin{equation}
\chi_{\bm k} \,\theta_{\bm k}-\sigma_{\bm k}^2=const\geqslant1/4
+x_{\bm k}^2\theta_{\bm k}+p_{\bm k}^2\chi_{\bm k}-( x_{\bm k}
p_{\bm k} +x_{-\bm k} p_{-\bm k})\sigma_{\bm k}, \label{inec}
\end{equation}
where $x_{\bm k}=<\phi_{\bm k}>$, $p_{\bm k}=<\pi_{\bm k}>$ are
the mean values of a scalar field and a momentum, correspondingly,
which satisfy to the equations of motion $x_{\bm k}^\prime=p_{\bm
k}/a^2$ and $p_{\bm k}^\prime=-a^2 k^2 x_{\bm k}$. Right hand side
of (\ref{inec}) is also integral of motion.

 The  inequality (\ref{inec})
represents the uncertainty relation \cite{hez}.

One can exclude $\chi_{\bm k}$ and $\theta_{\bm k}$ from Eqs.
(\ref{mean}) and obtain the single equation
\begin{equation}
 \sigma_{\bm k}^{\prime\prime\prime}=(4 k^2\sigma_{\bm k}+\sigma_{\bm k}^{\prime\prime})
 \left(\frac{a^{\prime\prime}}{a^\prime}
 -\frac{a^\prime}{a}\right)+4\left(\frac{{a^\prime}^2}{a^2}-k^2\right)\sigma_{\bm k}^\prime,
\label{single}
\end{equation}
where  $\chi_{\bm k}$ and $\theta_{\bm k}$ are expressed through
$\sigma_{\bm k}$:
\begin{eqnarray}
\chi_{\bm k}=-\frac{\sigma_{\bm k}^{\prime\prime}}{4 k^2 a^\prime
\,a}-\frac{\sigma_{\bm k}^\prime}{2 k^2a^2}
-\frac{\sigma_{\bm k}}{a^\prime a},\nonumber\\
\theta_{\bm k}=-\frac{\sigma_{\bm k}^{\prime\prime} a^3}{4
a^\prime}+\frac{1}{2}a^2 \sigma_{\bm k}^\prime -\frac{k^2 a^3
\sigma_{\bm k}}{a^\prime}.
\end{eqnarray}

For $x_{\bm k}=0$, $p_{\bm k}=0$, one can rewrite the uncertainty
relation in the form
\begin{equation}
(4 k^2\sigma_{\bm k}+\sigma_{\bm
k}^{\prime\prime})^2\frac{a^2}{{16 a^\prime}^2 k^2}-(4 k^2
\sigma_{\bm k}^2 +{\sigma_{\bm k}^\prime}^2)\frac{1}{4
k^2}\geqslant1/4 .
\end{equation}
Eqs. (\ref{single}) and (\ref{a}) have the additional integral of
motion:
\begin{equation}
M_p^2{a^\prime}^2+\frac{a}{2 a^\prime}\sum_{\bm k} 4 k^2
\sigma_{\bm k}+\sigma^{\prime\prime}_{\bm k}=const,
\end{equation}
which is the Friedman equation when the right-hand side is equal
to zero. However, the Friedman equation (Hamiltonian constraint)
is violated during a quantum epoch \cite{prep1,prep2}. This
violation, conserved at present leads to the non-zero right-hand
side constant.

\end{document}